# Personalize Web Searching Strategies Classification and Comparison

Mariya Savova Evtimova and Ivan Momtchilov Momtchev

*Faculty of Computer Systems and Control Programming and Computer Technologies Department,Technical University of Sofia, Sofia 1000, Bulgaria*

**Abstract**: Personalization is becoming very important direction in semantic web search for the users that needs to find appropriate information. In this paper, a classification of web personalization is proposed and semantic web search tools are investigated. Building user interest profile is essential for personalizing. Nowadays, semantic web tools use ontologies for personalization because of their advantages. It is important to mention that most of the semantic web search tools use agent technologies for implementation.

**Key words:** Web search engine, personalization, semantic web, user profile, information retrieval.

## 1. Introduction

The Web information increases the size constantly. Millions of internet pages are unorganized and finding relevant information is becoming difficult and need spreading certain amount of time [1, 2]. It is important for web developers and designers to know the behavior of the user, so that to be able to improve the quality of internet service. This way it is possible to predict what pages the user is interested in and to be presented to the user [3]. Today, for the web search, requirements increased, so that generating reasonable results from the search engine is not enough. Need for search engines to personalize returned results for each individual user, is becoming important [1, 2]. Definition of user interest profile to catch user intention is important in personalization.

In the paper is reviewed the classification and comparison of the personalized web search information and strategies for personalization is proposed. Section 2 presents the classification of the personalised web search information. In Section 3, is presented the web search tools details. Advantages of modern technics for personalization are discussed.

## 2. Classification of the Personalised Web Search Information

Technologies for personalization allow dynamic insertion, customization or proposition of the content of every format that is suitable for each individual user, concerning understanding behaviour of the user and his preference.

### 2.1 Web Pages Personalization

Web pages can be personalized based on characteristics like user interest, social categories, context of the individual user. Personalization define that the changes in the user preference is based on data as bought items and pages reviewed. Customization is used when web site use definite data like rating and preferences of the user. We can define three categories for personalization [3].
- Profile/Groups;
- Behavior;
- Collaboration.

### 2.2 Models for Web Personalization

Models for web personalization include rule filtering [4] based on principle of step by step

---

**Corresponding author:** Mariya Savova Evtimova, Ph.D., assistant professor, research fields: semantic web, object oriented programming, programming languages, distributed systems, Java technologies.

processing and collaborative filtering [5], that present relative material of the customers combining their own preference with the preference of others that in the same way. Collaborative filtering can applied for books, music, video and other. But is not applicable for other like apparel, jewellery, cosmetics and others. Another method "Prediction Based on Benefit" is applicable for products with complicated attributes such as apparel [3].

In the literature it is specified three methods for personalization [3, 6].

- Implicit;
- Explicit;
- Hybrid;

Implicit personalization is performed from the web pages or information system concerning different categories that of the personalization mentioned above. With explicit personalization the web page or information system is changed because of the functions provided from the system. Hybrid personalization combines both of the other two methods, so that to produce better results.

*2.3 Stages in Personalization*

The process of personalization can be divided into four stages [3, 7, 8].

2.3.1 Collection of Web Data

Implicit data include past events and click streams are stored in web server logs [9] or via cookies [9] or modules for tracing a sessions. Web logs provide enormous information for user action like IP address, time size for the requested URL and other. Using web logs, it is possible to identify user and predict future requests. That includes predict caching, beforehand extraction and representation. Historical data for access from the user visits of web servers are automatically stored in web access logs [8]. Web log data can be reviewed, as a collection of consistent access of the user, where each sequence is a list with pages accessible from one user from one session. Explicit data normally came from registration forms and rating questionnaires. Also, can be used demographic data and application data (e-transactions) [1, 5]. In some cases, web content, structure, and application data can be add as additional source of data in a subsequent processing.

2.3.2 Preprocessing of Web Data.

Data processing, usually is preprocessed to be put in format that is compatible with the technics for analysis that will be applied in the next stage. Preprocess can include cleaning from contradicted information, removing of unsuitable information for the aim for analysis (requests to integrated graphics will be added to server logs [9], although this adds little information for user interests), and completing of the missing links because of the caching in not entire path for clicking.

Web logs with requests are good source of information for the user behavior. Cluster query clustering algorithms are very famous for increasing efficiency of the web search. Cluster requests are useful for analysis of log data with requests that rely on user search history [9, 10].

Cleaning web log data—in different kind of files are stored all requests made from the user, some of them are suitable for navigation models for analysis of the data. Basic aim is to identify HTML documents and unnecessary files are removed in the following way [8]:

- A page requested from the user consists from other materials like graphics. They are removed, because they did not participate in process of analysis of the data;
- Sometimes user can request a page that does not exist;
- Some of the texts were not having URL addresses. That storage is erased, because URL addresses are mandatory during analysis of the navigation models.

Identification of the individual user—there is several ways to identify individual user. One way is to use cookies for inter-session coverage of the users or other using user registration that is not applicable for anonymous users. Basic problem is that IP address is not always identified with a single user. The task for defining of the user session is becoming more complicated with local cash, firewalls and proxy server. Missing of the cash hits in the server log makes complicated identification of the user process. Time spend of the user into one web page is very important characteristic for measuring user interest. Size of the file and the network traffic can affect real time spend in the web page.

Problems that can be observed in the web logs [8]:
- Identification of the users;
- Clients can have web access from different host;
- With proxy server one user can have many IP addresses;
- Missing of the data;
- Pages can be cashed;
- Ending session.

When additional and reliable analysis of visited web site is needed, the information from the log must be processed before use.

2.3.3 Analysis of Web Data.

Also known as Web Usage Mining this step includes machine learning technics and technics for analysis of the data for discovering user patterns and statistic correlation between web pages and user groups. This step normally usually leads to automatic user profiles and normally is applied offline, to avoid the load of the web server [3].

Research in the development of the web content concentrates on clusters, rules for association, semantic web, content analysis of the web page, analysis of search results, text analysis and image analysis. Data analysis covers customization, business intelligence, user's profiles, improving the system, recommendation, e-commerce, noise detection and web agents [7].

2.3.4 Final Stage/Taking Decision (Recommendation).

The last step of the process of personalization uses the results from the analysis of the data, to provide recommendation of the user. The process of recommendation usually include generating a dynamic web content, as adding a hyperlinks to the last requested page from the user. This can be realized by using carious technologies such as CGI programming. Personalization requires analysis of your goals and the development of business requirements, events and metrics. Integration of the tools for personalization is important, in order to better analyze the proposed information. Efficiency in finding relative information in internet depends on whether the specified queries to web search engine properly describe users information needs [9, 10].

## 3. Web Search Tools for Personalisation

Personalization is method that is used in variations of online behavior and individual differences that are observed in user's interests, information needs, goal for searching, contextual queries and other. Personalize system requires and maintains the information of the user such as demographic data, interests, preference and history of reviewed web pages [1].

There are several types of system for searching that provide information concerning individual needs. For example, hyperlink web searching, personalize web site and systems for recommendation [5].

Hyperlinks in internet are also important for personalization searching. Using personalized page rank to provide personalization web search was proposed in the beginning [5]. Problems that can be observed in web search engines in terms of individual users are:
- Ranking results is not appropriate for individual user but give good results to the audience;
- User queries are with low quality because the average length of the query is 2-3 words;
- Some words are poly-semantic and they have different meaning in different context;
- There is no possibility to implement a pattern for

**Table 1**

| Publication example [1] | Method | Description | Advantages |
|---|---|---|---|
| Personalized Semantic retrieval and summarization of web based documents [11] | Personalized search based on content analysis | User model with use interests | Use ontology to avoid cold start problem and information over load; semantic similarity |
| A ontological user modeling and semantic rule-based reasoning for personalization of help-on-demand services in pervasive environments [4] | Personalized search based on content analysis | Rule-based filtering | Use ontology to avoid cold start problem and information overload; highly expressive personalization |
| Bloom cookies: web search personalization without user tracking [12] | Personalized search based on content analysis | Profile generalization; noise injection | Privacy; personalization; network efficiency trade-off |
| Semantic relation based personalized ranking approach for engineering document retrieval [13] | Personalized search based on hyperlink analysis | Based on domain ontology relation-base ranking | Better than term-based ranking |
| Exploiting social bookmarking services to build clustered user interest profile for personalized search [14] | Personalized search based on hyperlink analysis | Page Rank | Ranking scores of the target URLs are better than google and bing search engine |
| Personalized web search by generating and mapping two user profiles [15] | Personalized search based on user group | Based on existing profile, two user profiles | Personalized web pages within short duration of time |
| Ensembling classifiers for detecting user's intentions behind web queries [10] | Personalized search based on user group | Ensemble of classifiers | Effectively detect user's intention |

searching because there is no possibility to adapt with the individual user.

Many attempts are made to personalize web search. Some methods use data for analysis, such as consumer e-mail, history of the visited websites, desktop files bookmark services that have low noise and provide accurate information.

Google Page Rank uses citation graph and link analysis. Google, Yahoo, and MSN are good for the experience user but for the other are not so effective. For individual user the browsing behavior should be observed.

Most of the semantic web search tools use software agents. Agents are useful when characterize social behavior and abilities that is important for reactiveness and contextualize process in web personalization [5]. Several strategies have been proposed concerning distinct problems, for instance, for semantically classification of search queries [10]. Search engine could be navigational tool to rich web site or can also provide access to different type of resources like books, music and other. Successful retrieval depends on automatically detecting user intention. This is known as supervised learning problem (classification) that uses word-based learning algorithms to search through a hypothesis space for appropriate hypothesis that is able to make good prediction, concerning intention detection.

Personalized web search strategies could be classified in three different groups: web search based on content analysis, hyperlink structure of the web and user groups. A review and comparison of personalized web search is proposed in Table 1.

## 4. Conclusions

In this paper classification and comparison for web search personalisation was presented. Analysis for the semantic technologies for personalisation is proposed. Nowadays, ontologies and agent technologies are widely used in semantic search tools for personalisation. Web search tools present different technologies concerning different direction of personalization. A review of different personalization methods and examples of the available web search tools are proposed.